\documentstyle[amssymb,aps,prl,epsfig,floats]{revtex}

\begin{document}
\draft

\twocolumn[\hsize\textwidth\columnwidth\hsize\csname@twocolumnfalse\endcsname

\title{Many-body $GW$ Calculations of Ground-State Properties:
Quasi-2D Electron Systems and van der Waals Forces}
\author{P.\ Garc\'{\i}a-Gonz\'alez$^1$ and R.~W.\ Godby$^2$}
\address{$^1$Departamento de F\'{\i}sica Fundamental, Universidad Nacional de
  Educaci\'on a Distancia,\\
Apartado 60141, 28080 Madrid, Spain}
\address{$^2$Department of Physics, University of York, Heslington, York YO10
  5DD, United Kingdom}
\date{\today}
\maketitle

\begin{abstract}
We present $GW$ many-body results for ground-state properties of two simple but
very distinct families of inhomogenous systems in which traditional 
implementations of density-functional theory (DFT) fail drastically. 
The $GW$ approach gives notably better results than the well-known random-phase 
approximation, at a similar computational cost. These results establish $GW$ as 
an superior alternative to standard DFT schemes without the expensive numerical 
effort required by quantum Monte Carlo simulations.
\end{abstract}

]

{\em Ab initio} many-body-theory (MBT) methods, particularly those based on
Hedin's {\em GW} approximation \cite{He65}, have been used extensively
during the last decade to calculate excited-state properties of electron
systems in solid-state physics \cite{AG98}. In addition, there is a recent
and increasing interest in the application of such methods to obtain
ground-state properties \cite{RG98,DW99,Ho99,ABL99,HA00,SFG00,PE01,GGG01}.
In MBT, electron-electron correlations are taken into account directly
without resorting to the mean-field density-based approximations used in
routine implementations of Density Functional Theory (DFT) \cite{HK64}, thus
providing a more microscopic description of the interacting many-electron
problem.

Modern MBT calculations use efficient algorithms for the evaluation of MBT
quantities [such as the dielectric function $\hat{\epsilon}\left( \omega
\right) $, the self-energy operator $\hat{\Sigma}\left( \omega \right) $,
and the one-particle Green function $\hat{G}\left( \omega \right) $]
required in the study of excited-state properties of real materials \cite
{AG94,ST95}. The application of these techniques to the calculation of
ground-state properties, such as the density or the total energy, opens the
appealing possibility of treating all the electronic properties in the same
fashion. Although more expensive than DFT, these MBT methods do not require
the demanding computational effort of quantum Monte Carlo simulations \cite
{QMC00}. Nonetheless, ground-state calculations based on MBT must be
painstakingly assessed. First, approximations that have proved successful
for spectral properties might not necessarily be good methods for structural
properties. Second, the energies and lifetimes of quasiparticles are mainly
determined by the pole structure of $\hat{G}\left( \omega \right) $, while
ground-state properties emerge from a multidimensional integration of $\hat{G%
}\left( \omega \right) $. As a consequence, features that can be safely
ignored in the determination of quasiparticle properties, like the
high-frequency behavior or high-energy-transfer matrix elements, play an
important role if we want to calculate, for instance, the ground-state
energy. The development of general-purpose procedures requires a careful
study of the optimal treatment of these points.

In this Letter we present nonselfconsistent (in the sense described below) 
{\em GW} calculations for the ground-state properties of two families of
simple inhomogeneous systems: the quasi-two-dimensional (2D) electron gas
and a pair of interacting jellium slabs. These provide a suitable test cases
for an extensive assessment of the performance of $GW$-MBT ground-state
calculations. First, despite the simplicity of these systems, LDA and GGA
fail to describe them properly. For the quasi-2D gas, the high inhomogeneity
of the density profile along the confining direction is clearly beyond the
scope of local or semilocal approaches \cite{KLN00}. The situation is quite
different if we consider interacting jellium slabs. In the limit of large
separation, in which the densities of each subsystem do not overlap,
dispersion or van der Waals (vdW) forces are evident. These forces are due
to long-ranged Coulomb correlations and, hence, cannot be described at all
by the LDA or GGA \cite{DW99,KMM98}. As shown below, even nonlocal DFT
prescriptions for the XC energy, such as the weighted density approximation
(WDA) \cite{GJL79}, are unable to reproduce such forces. On the other hand,
these systems exhibit translational invariance along the {\em xy} plane,
thus reducing significantly the number of independent variables needed to
describe the spatial dependence of all the operators, assisting the
production of the highly-converged test calculations required here.

In Hedin's $GW$ framework, the self-energy $\hat{\Sigma }$ of a system of $N$
electrons under an external potential $v_{{\rm ext}}\left( {\bf r}\right) $
is approximated by 
\begin{equation}
\Sigma \left( 1,2\right) =iG\left( 1,2^{+}\right) W\left( 1,2\right) ,
\label{1}
\end{equation}
where the labels 1 and 2 symbolize space-time coordinates. $\hat{W}$ is the
screened Coulomb potential, which is exactly related to the bare Coulomb
potential $\hat{w}$ and the polarizability $\hat{P}$ by $\hat{W}\left(
\omega \right) =\hat{w}+\hat{w}\hat{P}\left( \omega \right) \hat{W}\left(
\omega \right) =\hat{\epsilon }^{-1}\left( \omega \right) \hat{w}$ (the
usual matrix operations are implied). Under the $GW$ approach, the
polarizability is given by $P\left( 1,2\right) =-2iG\left( 1,2\right)
G\left( 2,1^{+}\right) $. Finally, $\hat{\Sigma }$ and $\hat{G}$ are linked
through the Dyson equation $\hat{G}^{-1}\left( \omega \right) =\omega - [ 
\hat{t}+\hat{v}_{{\rm el}}+\hat{\Sigma }\left( \omega \right)] $, with $\hat{%
t}$ the one-electron kinetic energy, and $v_{{\rm el}}\left( {\bf r}\right) $
the classical electrostatic potential [the sum of $v_{{\rm ext}}\left( {\bf r%
}\right) $ and the Hartree interaction potential]. This set of equations
defines a selfconsistent problem, but routine $GW$ calculations concerned
with the quasiparticle properties do not attempt selfconsistency: when
evaluating $\hat{\Sigma }$, the interacting Green function $\hat{G}$ is
generally substituted by that corresponding to the noninteracting Kohn-Sham
(KS) system under the LDA (or GGA) approximation.

It is well known that full selfconsistency implies a worsening of the
description of spectral properties \cite{ES98}. However, for the 3D \cite
{Ho99,GGG01} and 2D \cite{GGG01} homogeneous electron gases, the total
energy arising from $\hat{G}$ by using the Galitskii-Migdal formula of MBT 
\cite{GM58} fits the (essentially exact) QMC values extremely well if $\hat{G%
}$ is obtained selfconsistently. Nonetheless, at the nonselfconsistent level
(which we will call $G_{0}W_{0}$) the exact exchange energy is already built
in and, at the same time, correlation effects beyond the random-phase
approximation (RPA) are taken into account, as reflected in good total
energies and very good total energy differences \cite{Ho99,GGG01} for these
homogeneous systems. Since the relevant quantity for structural properties
is the total energy difference, $G_{0}W_{0}$ (and even, in some cases, RPA 
\cite{KP99}) is likely to fulfill a useful role. The known particle-number
violation under the $G_{0}W_{0}$ approximation \cite{Sc97}, while non-zero,
is so small that it can be ignored in practical applications \cite{SGG01}.

All the calculations reported in this Letter are done as follows. First, we
perform a standard LDA-KS calculation, in which, owing to the translational
invariance along the $xy$ plane, the KS orbitals are organized into {\em %
subbands}. That is, each KS state is given by $\phi _{n{\bf k}}\left( {\bf r}%
\right) =\psi _{n}\left( z\right) e^{i{\bf k}\cdot 
\mbox{\scriptsize
\boldmath $\rho$} }/ \sqrt{2\pi }$, where ${\bf k}=\left( k_{x},k_{y}\right) 
$ and $\mbox{\boldmath $\rho$} =\left(x,y\right) $ denote the
two-dimensional momentum and position in the $xy$-plane. Then, the
noninteracting polarizability $P_{0}\left( 1,2\right) =-2iG_{0}\left(
1,2\right) G_{0}\left( 2,1^{+}\right) $ is calculated at imaginary
frequencies using the expression 
\begin{equation}
P_{0}\left( z,z^{\prime },k;{\rm i}\omega \right) =\sum_{n=1}^{N_{{\rm occ}%
}}\sum_{m=1}^{\infty }S_{nmk}\left( {\rm i}\omega \right) \gamma _{nm}\left(
z\right) \gamma _{nm}\left( z^{\prime }\right)  \label{2}
\end{equation}
Here, $N_{{\rm occ}}$ is the number of occupied KS subbands, $\gamma
_{nm}\left( z\right) =\psi _{n}\left( z\right) \psi _{m}\left( z\right) $,
and the coefficients $S_{nmk}\left( {\rm i}\omega \right) $ admit a fully
analytical expression \cite{PE01,Eg85}. $P_{0}\left( {\rm i}\omega \right) $
could be also calculated fully numerically \cite{ST95} but due to the high
symmetry of the system, the present method is more efficient. The infinite
sum in Eq. \ref{2} is truncated at a certain value $N_{{\rm B}}$, which acts
as a convergence parameter.

The next step is the inversion of the RPA dielectric function $\hat{\epsilon}%
_{0}\left( {\rm i}\omega \right) =\hat{1}-\hat{w}\hat{P}_{0}\left( {\rm i}%
\omega \right) $ and the evaluation of $\hat{W}_{0}\left( {\rm i}\omega
\right) =\hat{\epsilon}_{0}^{-1}\left( {\rm i}\omega \right) \hat{w}$. This
can be done in a double cosine basis representation \cite{PE01,SGG01}, but
the high inhomogeneity of the systems along the $z$ direction suggests a
different method based on Ref. \onlinecite{AG94}. Note that the
polarizability spans a Hilbert space made up by the product states $\gamma
_{nm}$. Obviously, the set $\left\{ \gamma _{nm}\right\} $ has a large
number of linear dependences which can be eliminated using a standard
Gramm-Schmidt procedure, so obtaining an optimized basis set ${\cal B}_{{\rm %
op}}=\left\{ \zeta _{\alpha }\left( z\right) \right\} $. In all the cases
here studied, $N_{{\rm op}}=N_{{\rm B}}+N_{{\rm occ}}-1$ $\zeta $-functions
ensure an almost perfect representation (with a relative error less than $%
10^{-4}$) of all the product states $\gamma _{nm}$. Hence, $P_{0}\left(
z,z^{\prime },k;{\rm i}\omega \right) =\sum_{\alpha ,\beta }\zeta _{\alpha
}\left( z\right) P_{0}\left( k,{\rm i}\omega \right) _{\alpha \beta }\zeta
_{\beta }\left( z^{\prime }\right) $, where the matrix elements $P_{0}\left(
k,{\rm i}\omega \right) _{\alpha \beta }$ can be immediately obtained from $%
S_{nmk}$ and the scalar products $\left\langle \gamma _{nm},\zeta _{\alpha
}\right\rangle $. Then, we calculate the representation $w\left( k\right)
_{\alpha \beta }$ of the bare Coulomb potential, and the matrix elements $%
W\left( k,{\rm i}\omega \right) _{\alpha \beta }$ of the screened Coulomb
potential are easily obtained by a matrix inversion for each value of $k$.
Well converged results are typically obtained with $N_{{\rm B}}\simeq 80$
subbands, although for quasi-2D systems $N_{{\rm B}}$ is significantly less.
RPA correlation energies, where required, can be straightforwardly obtained
in this representation.

The real-space representation of $\hat{W}_{0}$ is given by the expansion 
\begin{eqnarray*}
W_{0}\left( z,z^{\prime },;{\rm i}\tau \right)  &=&{\rm i}\sum_{\alpha
,\beta ,{\bf k}}\int \frac{d\omega }{2\pi }e^{{\rm i}\left( \omega \tau +%
{\bf k}\cdot \mbox{\scriptsize \boldmath $\rho$}\right) } \\
&&\times \zeta _{\alpha }\left( z\right) \zeta _{\beta }\left( z^{\prime
}\right) W\left( k,{\rm i}\omega \right) _{\alpha \beta }\;,
\end{eqnarray*}
where $\sum_{{\bf k}}=\left( 2\pi \right) ^{-2}\int d{\bf k}$, whereas the
KS Green function $G_{0}\left( z,z^{\prime },\rho ;{\rm i}\tau \right) $ is
calculated directly from the KS eigenstates \cite{ST95}. By using (\ref{1})
we readily get the self-energy operator in real space and imaginary time
and, eventually, its representation $\Sigma \left( k,{\rm i}\tau \right)
_{nm}$ in the KS basis set. We use typically 100 (or fewer) $\tau /\omega $
points in a Gauss-Legendre grid with $\omega _{\max }\simeq 80\Delta _{c}$ ($%
\Delta _{c}$ is the width of the conduction band), and 100-150 $\rho /k$
points with $k_{\max }^{2}/2\simeq \omega _{\max }$. It is important to
emphasize that the asymptotic time and frequency tails of all the operators
must be treated analytically to ensure smooth and rapid convergence.

In the imaginary-time/frequency representation $\hat{\Sigma }\left( \mu +%
{\rm i}\omega \right) =-{\rm i}\int d\tau \,\hat{\Sigma }\left( {\rm i}\tau
\right) e^{-{\rm i}\omega \tau }$, with $\mu $ the interacting chemical
potential. Hence, the Green function at imaginary frequencies is the
solution of the Dyson equation 
\begin{equation}
\hat{G}^{-1}\left( \mu +{\rm i}\omega \right) =\mu +{\rm i}\omega -\left[ 
\hat{h}_{{\rm KS}}+\hat{\Sigma }\left( \mu +{\rm i}\omega \right) -\hat{v}_{%
{\rm XC}}\right]  \label{4}
\end{equation}
with $\hat{h}_{{\rm KS}}=\hat{t}+\hat{v}_{{\rm S}}$ the KS-LDA hamiltonian
and $v_{{\rm XC}}\left( {\bf r}\right) $ the LDA-XC potential. $\hat{G}$ is
calculated in the KS representation $G\left( k,{\rm i}\omega \right) _{nm}$,
with $\mu $ previously obtained by diagonalizing the quasiparticle
hamiltonian at ${\rm i}\omega =0$, $\hat{h}_{{\rm QP}}\left( \mu \right) =%
\hat{h}_{{\rm KS}}+\hat{\Sigma }\left( \mu \right) -\hat{v}_{{\rm XC}}$, and
by imposing that the volume enclosed by the interacting Fermi surface equals
the KS value. A small term $\delta v\left( {\bf r}\right) $, accounting for
the change in the Hartree potential due to the differences between the $%
G_{0}W_{0}$ and LDA densities, should be included into (\ref{4}) and $\hat{h}%
_{{\rm QP}}$. However, this term induces an imperceptible change in the
shape of $\hat{G}$, and it has a negligible influence on the ground-state
properties.

\begin{figure}[t!]
\epsfxsize=8cm \centerline{\epsfbox{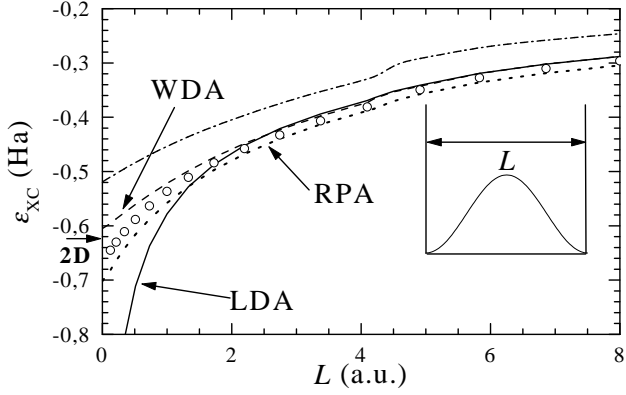}} \bigskip
\caption{XC energy per particle for thin confined jellium slabs of fixed 2D
density $n^{{\rm 2D}} = 3/4\pi$ as a function of their thickness $L$. Lines:
LDA, WDA, and RPA; circles: $G_0W_0$. The exchange energy (dash-dotted line)
has been included as a reference. Note the effect due to the filling of a
second subband at $L \simeq 4.2a_0$, which is especially evident in $%
\varepsilon_{{\rm X}}$. The arrow indicates the QMC value for the 2D limit.}
\label{Fig1}
\end{figure}

The electron density and energy are given, respectively, by $n\left( {\bf r}%
\right) =\pi ^{-1}\oint d\omega \,G\left( {\bf r,r};{\rm i}\omega \right) $
and $E=E_{{\rm el}}+\pi ^{-1}\,{\rm tr}\oint d\omega [\hat{t}+\frac{1}{2}%
\hat{\Sigma}\left( {\rm i}\omega \right) ]\hat{G}\left( {\rm i}\omega
\right) $, where $E_{{\rm el}}$ is the classical electrostatic energy [which
can be obtained from $n\left( {\bf r}\right) $], and tr symbols the spatial
trace. Owing to the symmetry of our model systems (the generalization to
arbitrary geometries is straightforward), we can write $n\left( z\right)
=n_{0}\left( z\right) +\delta n\left( z\right) =n_{0}\left( z\right)
+\sum_{n,{\bf k}}\left| \psi _{n}\left( z\right) \right| ^{2}\delta f_{nk}$, 
$n_{0}\left( z\right) $ being the LDA density, and 
\begin{equation}
\delta f_{nk}=\int \frac{d\omega }{\pi }\left[ G\left( k,\mu +{\rm i}\omega
\right) _{nn}-G_{0}\left( k,\mu _{0}+{\rm i}\omega \right) _{nn}\right] \,,
\label{4.b}
\end{equation}
($\mu _{0}$ is the LDA chemical potential). On the other hand, the exchange
energy per surface unit is given directly by $E_{{\rm X}}/S=\frac{1}{2}%
\sum_{n,{\bf k}}\,f_{nk}\Sigma _{{\rm X}}\left( k\right) _{nn}$, where $\hat{%
\Sigma}_{{\rm X}}$ is the frequency-independent part of the self-energy $%
\Sigma _{{\rm X}}\left( {\bf r}_{1},{\bf r}_{2}\right) ={\rm i}\lim_{\tau
\rightarrow 0^{+}}G_{0}\left( {\bf r}_{1},{\bf r}_{2};{\rm i}\tau \right)
w\left( r_{12}\right) $ (i.e., the Fock operator). Finally, the correlation
energy $E_{{\rm C}}$ is written as the sum of its kinetic-energy and
electrostatic parts: 
\begin{eqnarray}
\frac{T_{{\rm C}}}{S} &=&\sum_{n,{\bf k}}\varepsilon _{nk}\delta f_{nk}-\int
dzv_{{\rm S}}\left( z\right) \delta n\left( z\right)   \label{6} \\
\frac{W_{{\rm C}}}{S} &=&\sum_{n,{\bf k}}\left[ \frac{1}{2}\Sigma _{{\rm X}%
}\left( k\right) _{nn}\delta f_{nk}+\right.   \nonumber \\
&&\left. \sum_{m}\int \frac{d\omega }{\pi }\Sigma _{{\rm C}}\left( k,\mu
+i\omega \right) _{nm}G\left( k,\mu +i\omega \right) _{mn}\right]   \label{7}
\end{eqnarray}
where we have included the frequency-dependent part of the self-energy $\hat{%
\Sigma}_{{\rm C}}\left( {\rm i}\omega \right) =\hat{\Sigma}\left( {\rm i}%
\omega \right) -\hat{\Sigma}_{{\rm X}}$. As usual, we suppose that the total
electrostatic energy is given correctly by the LDA, since we have checked
carefully that the density variation $\delta n\left( z\right) $ causes only
minor changes in the electronic energy. We also note that when evaluating $%
E_{{\rm C}}$ we do not need all the matrix elements of $\hat{\Sigma}$ and $%
\hat{G}$. In general, full convergence is achieved by performing the sum
only over LDA states such that $\varepsilon _{n{\bf k}}\lesssim 6\Delta _{c}$%
, and this also applies to the resolution of Dyson's equation. Therefore,
the most demanding part of the calculation is the evaluation of $\hat{%
\epsilon}^{-1}\left( {\rm i}\omega \right) $ which, in any case, is required
to obtain the RPA correlation energy.

To mimic a quasi-2D electron system, we have taken a thin jellium slab with
a background density $\overline{n}=(\frac{4}{3}\pi r_{{\rm s}}^{3})^{-1}$
and width $L$. This slab is bounded by two infinite planar walls, and
overall charge neutrality is assumed. We keep the number of particles per
unit surface area $n_{{\rm 2D}}=\overline{n}L$ constant, in such a way that
the limit $L\rightarrow 0$ corresponds to a 2D homogeneous electron gas
(HEG) with density $n_{{\rm 2D}}$. In Fig. 1 we plot, for several values of $%
L$, the XC energy per particle $\varepsilon _{{\rm XC}}$ given by the LDA,
the nonlocal WDA, the RPA, and the $G_{0}W_{0}$ method. As commented
previously, the LDA diverges when approaching the 2D limit, whereas the WDA
behaves reasonably well, slightly underestimating the absolute value of $%
\varepsilon _{{\rm XC}}$ in the strict 2D limit. The RPA does not show any
pathological behaviour, but it overestimates $\left| \varepsilon _{{\rm XC}%
}\right| $ by more than 20 mHa/e$^{-}$ for all configurations. Finally, $%
G_{0}W_{0}$, whose superiority to the RPA has already been established in
the 3D limit,\cite{Ho99,GGG01} retains this in the transition to the 2D
limit. Its performance is similar to that of the WDA for these systems
(although the residual error for the 2D gas has the opposite sign). It is
important to point out that the RPA and $G_{0}W_{0}$ XC energies obtained by
starting from a WDA-KS calculation are indistinguishable from those plotted
in Fig. 1. Thus, the specific choice of the KS method seems to be of minor
importance when calculating XC energies using MBT.

The study of the interacting energy between two unconfined jellium slabs is
of more direct significance, as it has been considered as a benchmark for 
{\em seamless} correlation functionals attempting to describe vdW forces 
\cite{DW99,Ry00}. By varying the distance $d$ between the slabs we cover
configurations in which the density profiles of each subsystem overlap (i.e.
there is a {\em covalent} bond), and situations (when $d\gg 0$) in which
there is no such overlap and the only source of bonding is the appearance of
vdW forces. In the upper panel of Fig. 2 we plot the XC energy per particle $%
\varepsilon _{{\rm XC}}\left( d\right) $ as a function of $d$ using the LDA
(the WDA gives very similar results), the RPA, and the $G_{0}W_{0}$, for two
slabs of width $L=12a_{0}$ and a background density $\overline{n}$
corresponding to $r_{{\rm s}}=3.93$. We can see that the $G_{0}W_{0}$
reduces the RPA error by 60\%. In the lower panel we represent the
correlation binding energy per particle [defined as $e_{{\rm C}}\left(
d\right) =\varepsilon _{{\rm C}}\left( d\right) -\varepsilon _{{\rm C}%
}\left( \infty \right) $]. First, neither the LDA nor even the nonlocal WDA
is able to reproduce the characteristic asymptotic $d^{-2}$ vdW behavior.
This behavior {\it is} described by the\ RPA and the $G_{0}W_{0}$, and for
both approximations $e_{{\rm C}}\left( d\gg 0\right) $ is very similar
(which is not a surprise because such an asymptotic behavior is fully
described at the RPA level \cite{DW99}). For intermediate and small
separations, there are slight differences between the RPA and the $G_{0}W_{0}
$ but much less important than those appearing when comparing the total
correlation energies.

A detail that is worth pointing out is the fact that the error in the {\em %
absolute} $G_{0}W_{0}$ correlation energy is amenable to a LDA-like
correction. From the differences, given in Ref. \onlinecite{GGG01}, between
QMC and $G_{0}W_{0}$ correlation energies for the HEG, we can build a
functional $\Delta E_{{\rm C}}=\int d{\bf r}n\left( {\bf r}\right) \,\delta
\varepsilon _{{\rm C}}\left[ n\left( {\bf r}\right) \right] $. The absolute $%
G_{0}W_{0}$ energy corrected in this way is in broad correspondence with the
LDA\ energy, while the binding energy retains its correct $1/d^{2}$ behavior
at large $d$ and is little altered for small $d$ (see Fig. 2).

\begin{figure}[t!]
\epsfxsize=8cm \centerline{\epsfbox{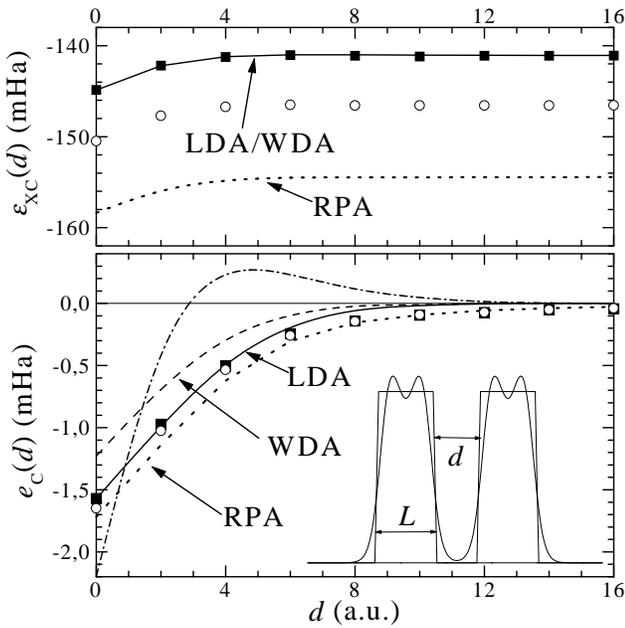}} \bigskip
\caption{Upper panel: XC energy per particle for two jellium slabs as a
function of the distance $d$. Lines: LDA, WDA and RPA; circles: $G_0W_0$;
squares: $G_0W_0$ plus LDA-like correction term. Lower panel: correlation
binding energy per particle. The LDA and WDA contributions have been
obtained by subtracting the exact exchange energy from the corresponding XC
values. The exchange binding energy (dashed-dotted line) has been also
included in this panel.}
\label{Fig2}
\end{figure}

In summary, we have performed $GW$-MBT calculations to evaluate the
ground-state energy of inhomogeneous systems. With practically the same cost
we have obtained correlation energies beyond the random phase approximation.
The importance of a truly {\em ab initio} treatment of electron many-body
effects is evident in the model systems we have chosen, for which
traditional implementations of DFT are completely inaccurate.

The authors thank C.-O.\ Almbladh, J.E. Alvarellos, U. von Barth, E.
Chac\'{o}n, J. Dobson, B. Holm, P. Rinke, and A. Schindlmayr for many
valuable discussions. This work was funded in part by the EU through the
NANOPHASE Research Training Network (Contract No.\ HPRN-CT-2000-00167) and
by the Spanish Education Ministry DGESIC grant PB97-1223-C02-02.

\end{document}